\documentclass[11pt,a4paper]{article}
\usepackage[margin=1in]{geometry}
\usepackage[T1]{fontenc}
\usepackage{lmodern}
\usepackage[numbers,square,sort&compress]{natbib}
\usepackage{amsmath,amssymb,booktabs,array,microtype,graphicx}
\usepackage{placeins}
\usepackage{hyperref}
\usepackage{url}
\hypersetup{hidelinks,
 pdftitle={Post-training weight repair for simulation-based inference},
 pdfauthor={Benedikt Maier and Michael Spannowsky}}
\newcommand{\E}{\mathbb{E}}
\newcommand{\Var}{\operatorname{Var}}

\newcommand{\softplus}{\operatorname{softplus}}
\newcommand{\vefr}{\textsc{VEFR}}
\newcommand{\dd}{\mathrm{d}}

\title{Post-training weight repair for simulation-based inference}
\author{%
Benedikt Maier\textsuperscript{1}\quad Michael Spannowsky\textsuperscript{2,3}\\[0.7em]
\small\textsuperscript{1}Department of Physics, Imperial College London, London, United Kingdom\\
\small\textsuperscript{2}Institute for Theoretical Physics, Karlsruhe Institute of Technology,\\
\small Karlsruhe, Germany\\
\small\textsuperscript{3}Institute for Quantum Materials and Technologies (IQMT),\\
\small Karlsruhe Institute of Technology, Germany}
\date{}
\begin{document}
\maketitle

\begin{abstract}
For a given choice of model parameters, we can often simulate what an experiment might observe. Interpreting the data takes us in the opposite direction, from the observations to the parameter values that best describe them. When the likelihood cannot be calculated directly, one approach to this inverse problem is to train a neural network on simulated events to estimate likelihood ratios. This form of simulation-based inference can require a large and costly training sample. Even a small error that repeatedly favours one parameter value can shift the best fit when many events are combined.

We introduce verified evolutionary functional repair to improve the accuracy of the fit using simulations already in hand. Gradients of the training loss guide the search for changes to the network weights. To decide whether to keep a change, we fit the same mock datasets with the original and edited networks. Both see the same statistical fluctuations, so we can compare their errors directly.

In a detector-level study of a CP-odd Higgs coupling, all 18 networks gave more accurate coupling estimates after repair. After fixing the edits, we measured the gain on new mock datasets drawn from the same simulated events. Because the parameters used to generate a simulated dataset are known, other analyses can test how accurately their full fit recovers them even when the exact likelihood is inaccessible. The improvement at a fixed simulation budget suggests that weight repair could help an analysis reach its target accuracy with fewer simulated training events.
\end{abstract}

\section{Introduction}
Simulation-based inference (SBI) uses data simulated at known parameter values to infer model parameters from observations when the likelihood cannot be evaluated directly~\citep{cranmer2020frontier}. A neural likelihood estimator can require a large training sample, and the accuracy attainable with a given number of simulations depends on both the method and the inference problem~\citep{lueckmann2021benchmarking}. When further events are expensive to simulate, there is good reason to look again at the trained estimator and ask which of its errors matter for the fit.

Consider a network that estimates, for each event, the likelihood ratio between two values of a parameter~\citep{hermans2020likelihood,durkan2020contrastive}. Suppose it slightly overstates the evidence for the larger value. The error may be hard to see in any one event, yet the fit adds the log likelihood ratios from all the independent events in the dataset. If the same preference recurs, it accumulates in the sum and can shift the best fit towards larger values. In simulation we know the parameter value used to generate the events, so repeated fits to mock datasets can reveal a shift that the average training loss gives us little reason to suspect.

These repeated fits are the basis of verified evolutionary functional repair (\vefr), our method for improving a likelihood estimator after training. We use the gradient of the training loss to identify promising changes to the network weights. To test a change, we assemble mock datasets, or pseudoexperiments, from the existing simulations and fit each one with both the original and the edited network. Both estimators then see the same statistical fluctuations. We keep an edit only if the paired fits show that it improves recovery of the generating parameters.

Much of the work on reducing the simulation cost of SBI concerns how to obtain the estimator in the first place. Sequential inference~\citep{papamakarios2016fast,lueckmann2017flexible,papamakarios2019sequential,greenberg2019automatic}, Bayesian optimisation~\citep{gutmann2016bayesian} and prior truncation~\citep{miller2021truncated} use earlier simulations to decide where in parameter space to generate the next sample. During training, additional labels extracted from the simulator can make each event more informative~\citep{brehmer2020mining}, and an estimator pretrained on cheaper simulations can be fine-tuned on more realistic ones~\citep{saoulis2025transfer}. Weight repair enters after these choices have been made, when a trained estimator and simulated samples are already available.

Letting the intended measurement guide the design of an estimator has precedents in SBI. Summary statistics can be chosen for the particular parameter one wishes to measure~\citep{fearnhead2012constructing}. When nuisance parameters affect the observations, one can train a statistic for precision in the fit~\citep{castro2019inferno,heinrich2022learning} or train it to be insensitive to nuisance variations~\citep{louppe2017learning}. In \vefr{} the measurement guides a choice that remains open after training, namely which changes to the likelihood estimator improve parameter recovery.

Recent studies have used frontier language models to edit network weights directly~\citep{takehi2026weightspace} or to write programs that update weights in response to experimental feedback~\citep{li2026weightloop}. An SBI analysis already supplies a test of any proposed edit because the generating parameters are known. In \vefr{} the search proposes weight changes without a language model or an agent, and the physics fit determines which change to keep. Other approaches train networks for later correction~\citep{sinitsin2020editable}, learn from the weights of a trained network~\citep{navon2023equivariant}, or edit language models through learned gradient transformations or rank-one updates~\citep{mitchell2022mend,meng2022locating}. For an estimator used in a measurement, the same paired fits could test weight changes proposed by another editing method.

We illustrate \vefr{} by estimating a CP-odd coupling in electroweak Higgs production with two jets. The coupling changes the reconstructed kinematics, but a shift in the jet energy scale also affects some of the same distributions. A useful repair must therefore improve the coupling estimate when the fit allows for uncertainty in the jet calibration. After fixing the edits, we evaluated each network twice on new mock datasets assembled from the existing simulations. All 18 networks gave a lower root-mean-square error (RMSE) in the fitted coupling after repair in both runs.

Moving the best fit closer to the true coupling leaves open whether the quoted confidence set contains that coupling as often as it should. In Bayesian SBI, learned posteriors can give overconfident credible sets~\citep{hermans2022trust}. Approaches to this problem include balancing the training of a ratio estimator~\citep{delaunoy2022balanced}, incorporating coverage into the training loss~\citep{falkiewicz2023calibrating} and calibrating credible sets after training~\citep{cabezas2025cp4sbi}. In our frequentist fit we assess coverage separately from the error used to choose the repair. At each generating parameter setting we measure how often the confidence set contains the true coupling over repeated mock experiments.

Section~\ref{sec:vefr} gives the procedure for applying \vefr{} to an SBI estimator. Sections~\ref{sec:statistical} and~\ref{sec:collider} describe the coupling fit and the simulated events, and Section~\ref{sec:method} gives the search settings used in this application. Section~\ref{sec:results} reports the tests of the fixed repairs and examines how sensitive the gain is to the original sample of simulated events.

\section{Verified evolutionary functional repair}
\label{sec:vefr}

\begin{figure}[!tb]
\centering
\includegraphics[width=\textwidth]{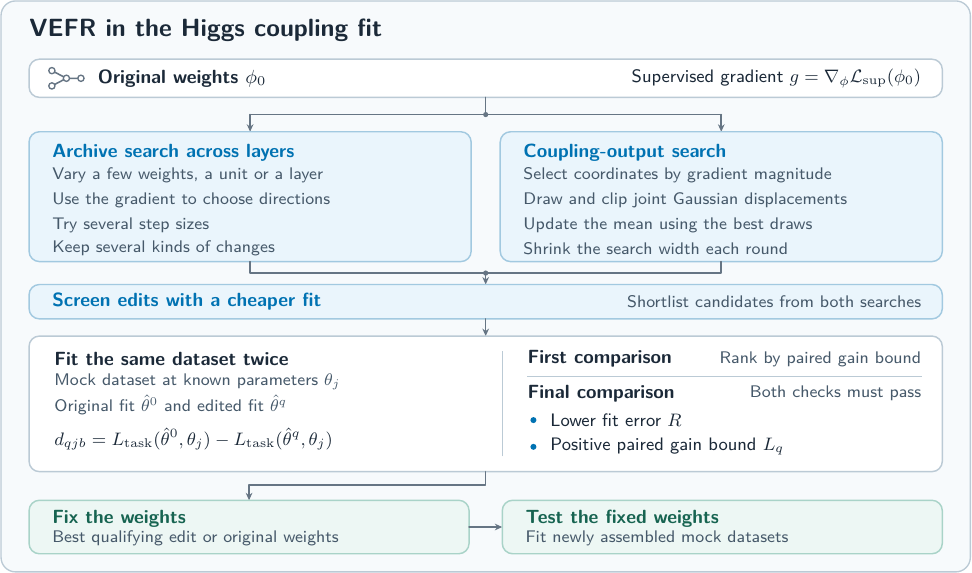}
\caption{VEFR as implemented for the Higgs coupling fit. This benchmark uses two proposal searches and two rounds of paired fits. One search varies weights across the network, and the other makes joint changes to selected weights and biases of the coupling outputs. The first round of fits ranks the shortlisted edits. The final round tests the leading candidates against the original estimator before the selected weights are fixed for evaluation.}
\label{fig:repair-overview}
\end{figure}

A trained estimator with weights $\phi_0$ leaves us considerable freedom to change how it responds to an event. The architecture determines which weight changes are possible, but it is the intended measurement that gives us a reason to prefer one change over another. In \vefr{}, labelled simulations guide the search for promising changes. For the shortlisted candidates, we apply the intended fit to mock datasets and compare the estimates with the parameter values used to generate them. Figure~\ref{fig:repair-overview} shows how we use this procedure in the Higgs application.

\subsection{Preparing an analysis for repair}

A change that improves a fit near one parameter value may be less useful elsewhere. We therefore choose generating points $\theta_1,\ldots,\theta_J$ across the region of interest. The size of the dataset matters too. In a collision measurement a small systematic error can become important when many events are combined, so we need enough simulated events to assemble mock datasets of the size used in the analysis. The mock experiments also include the relevant nuisance parameter settings and any auxiliary measurements that constrain them in the fit.

For a mass measurement, each mock experiment gives us a fitted mass to compare with the mass used to generate it. Their squared difference is one choice for the loss $L_{\rm task}$ of a single fit. To express the accuracy of the measurement in mass units, we take the square root of the mean squared error over mock experiments at each generating mass. Averaging these root-mean-square errors (RMSEs) over the chosen masses gives the performance measure $R$. We use the analogous definitions for the Higgs coupling. Other measurements can use different measures of error, with smaller values of both $L_{\rm task}$ and $R$ indicating better fits.

An edit that looks promising on the events that suggested it may owe some of its apparent success to fluctuations in those events. We therefore set aside separate samples before the search for computing edit directions, screening proposals and assembling the mock datasets for the paired fits. The first sample carries the labels used by the supervised training loss, such as class labels for a ratio estimator trained by classification. Different reconstructions of the same simulated event carry some of the same information, so they stay together when we divide the samples. If likelihood ratios are normalised on a finite reference sample, changing the network can also change its normalisation. We recompute it for each candidate on the same reference events so that a different draw of events cannot favour one candidate over another.

\subsection{Proposing and screening weight changes}

Even a modest network offers many more possible weight changes than we can test in the full fit. The gradient of the supervised loss at $\phi_0$ gives us a place to start. The components with the largest absolute values identify the weights to which that loss is locally most sensitive. One proposal is a step along the negative gradient restricted to a few such weights. Another uses the leading singular vectors of a layer's gradient matrix to change many weights together. We try several step sizes and groups of weights suited to the architecture. In the Higgs application, a second search varies a small group of weights jointly, using the best-scoring draws to update the distribution for the next round. In both cases a proposal is a displacement $\Delta\phi_q$ from the original weights, giving the candidate $\phi_q=\phi_0+\Delta\phi_q$.

Running the full fit for every proposal would be expensive, so we first screen the candidates with a cheaper fit. This fit may use smaller mock datasets or hold a nuisance parameter fixed. An edit that ranks highly when the nuisance parameter is fixed can lose that advantage once the parameter is free to vary. We therefore keep promising edits of several kinds on the shortlist, for example changes to a hidden layer and changes confined to the outputs, so that the full fit can decide between them. Repeating the search requires the allowed weight groups, step sizes or sampling distributions, shortlisting rules and random seeds.

\begin{samepage}
\subsection{Choosing an edit by paired fits}
\label{sec:paired-repair}

A single mock experiment can favour an edit by chance. We therefore require the mean gain in task loss to remain positive after allowing for fluctuations across mock experiments. We also require the error measure $R$ chosen for the measurement to fall. In the mass example these are distinct checks. A reduction in squared error at a mass where the original fit was poor can dominate the mean gain, even if the RMSE in mass units, averaged over the generating masses, gets worse.

To measure the gain for candidate $q$, we fit each mock dataset $D_{jb}$ twice, once with the original weights and once with the edit. Here $j$ indexes the generating parameter point and $b$ the draw. Both fits see exactly the same data, including any auxiliary measurements, so the fluctuations in the inputs are common to the two estimates $\widehat\theta^{\,0}_{jb}$ and $\widehat\theta^{\,q}_{jb}$. The reduction in loss on this dataset is
\begin{equation}
 d_{qjb}=L_{\rm task}(\widehat\theta^{\,0}_{jb},\theta_j)
         - L_{\rm task}(\widehat\theta^{\,q}_{jb},\theta_j).
\end{equation}
\par
\end{samepage}
Suppose a dataset generated at a mass of $100\,\mathrm{GeV}$ gives a fitted mass of $104\,\mathrm{GeV}$ with the original weights and $102\,\mathrm{GeV}$ after the edit. The positive gain, $(4^2-2^2)\,\mathrm{GeV}^2=12\,\mathrm{GeV}^2$, tells us that the edited estimate is closer for this dataset. Repeating the comparison on $E_j\geq 2$ mock datasets at each generating point gives the mean gain $\overline d_{qj}$ and the sample variance of the individual paired gains $\widehat{\Var}_b(d_{qjb})$. We give the generating points equal weight. For independent draws conditional on the simulated event samples, the overall mean gain, its estimated Monte Carlo standard error and the lower endpoint of an approximate two-sided $95\%$ interval are
\begin{equation}
 \widehat\Delta_q=\frac1J\sum_j\overline d_{qj},\qquad
 \widehat s_q=\frac1J
 \left[\sum_j\frac{\widehat{\Var}_b(d_{qjb})}{E_j}\right]^{1/2},
 \qquad L_q=\widehat\Delta_q-1.959964\,\widehat s_q .
 \label{eq:lcb}
\end{equation}
The two requirements are thus $L_q>0$ and $R(\phi_q)<R(\phi_0)$. To follow the mass example through to this decision, suppose the original fits have RMSEs of $4$ and $3\,\mathrm{GeV}$ at generating masses of $100$ and $120\,\mathrm{GeV}$, and the edited fits have RMSEs of $2\,\mathrm{GeV}$ at both masses. The mean paired gains are $4^2-2^2=12$ and $3^2-2^2=5$ in $\mathrm{GeV}^2$, giving $\widehat\Delta_q=8.5\,\mathrm{GeV}^2$. If the standard errors of the two mean gains are $2$ and $1\,\mathrm{GeV}^2$, the standard error of the average gain is $\sqrt{2^2+1^2}/2\simeq1.12\,\mathrm{GeV}^2$. The lower endpoint is then $L_q\simeq6.31\,\mathrm{GeV}^2$, above zero. The mean RMSE also falls from $3.5$ to $2\,\mathrm{GeV}$, so the edit qualifies.

Among qualifying edits we choose the largest $L_q$, breaking ties by the larger reduction in $R$ and then by candidate identifier. If none qualifies, we keep the original weights. This paired check is what we mean by verification. The interval applies to a fixed candidate conditional on the event samples. Searching many candidates can favour one with an unusually large gain on the selection draws~\citep{cawley2010overfitting,dwork2015reusable}. We therefore fix the selected weights and repeat the paired fits on newly assembled mock datasets to measure the improvement after selection.

\section{Parameter inference with learned likelihood ratios}
\label{sec:statistical}

In the collider benchmark the parameter of interest, $c$, sets the strength of a CP-odd contribution to Higgs production. Changing $c$ changes the distribution of the two jets produced with the Higgs boson. The energy calibration also affects this distribution. A positive fractional shift $\eta$ in the jet energy scale raises the reconstructed jet momenta and can move events across the selection threshold. To allow for this uncertainty, we fit $c$ and $\eta$ together, using a separate measurement $a$ of the fractional scale shift to constrain $\eta$. The nominal point $(c,\eta)=(0,0)$ corresponds to Standard Model-like production and the nominal jet calibration.

\subsection{Building the inference problem from event likelihoods}

Let $x$ denote the measured observables of a selected event and $\theta=(c,\eta)$ the trial parameters. We compare hypotheses using a common reference density $p_{\rm ref}(x)=p(x\mid 0,0)$. For a given event this reference factor cancels when comparing two hypotheses. A network with weights $\phi$ estimates the ratio of the two selected-event densities,
\begin{equation}
 r_\phi(x;\theta)\simeq \frac{p(x\mid\theta)}{p_{\rm ref}(x)}.
\end{equation}
The reference must have support wherever a trial density is positive. Under this condition, averaging the exact ratio over reference events gives $\int p(x\mid\theta)\,\dd x=1$. We impose the same normalisation on an independently simulated reference sample $\mathcal Z$ by dividing the positive network output $\widetilde r_\phi$ by its weighted sample mean. The positive importance weights $\omega_i$ target the reference density, giving
\begin{equation}
 r_\phi(x;\theta)=\frac{\widetilde r_\phi(x;\theta)}
 {\widehat Z_\phi(\theta)},\qquad
 \widehat Z_\phi(\theta)=
 \frac{\sum_{i\in\mathcal Z}\omega_i\widetilde r_\phi(x_i;\theta)}
 {\sum_{i\in\mathcal Z}\omega_i}.
 \label{eq:normalizer}
\end{equation}
The normaliser contributes $-n\log\widehat Z_\phi(\theta)$ to the log likelihood for $n$ events. Its dependence on $\theta$ therefore matters to the fit. We recompute the normaliser for every edit and parameter value using the same reference events. A finite sample gives an approximate normalisation under $p_{\rm ref}$, and using the same events correlates the normalisation errors across the comparisons.

We condition on $n=1000$ selected events to isolate the information in their kinematic distributions. At fixed $\theta$, we model the events as independent draws and the auxiliary measurement as an independent Gaussian observation with mean $\eta$ and standard deviation $\sigma_a$. Their log likelihood contributions add, giving the approximate log likelihood up to a parameter-independent constant,
\begin{equation}
 \ell_\phi(c,\eta)=\sum_{i=1}^{n}\log r_\phi(x_i;c,\eta)
 -\frac{(a-\eta)^2}{2\sigma_a^2},\qquad \sigma_a=0.02.
 \label{eq:likelihood}
\end{equation}
At the generating jet energy scale we draw $n$ selected events with replacement. Generator event weights set the sampling probabilities at each coupling, allowing the same simulated collisions to represent several values of $c$. The weights that \vefr{} changes are the network parameters $\phi$ that determine the predicted likelihood ratio. Both networks fit the same events and the same auxiliary draw $a\sim\mathcal N(\eta_{\rm true},\sigma_a^2)$. For each trial coupling we maximise the likelihood over $\eta\in[-0.05,0.05]$, allowing the scale to adjust within the fit. The best-fit coupling then maximises this profile likelihood over $c\in[-1,1]$, and the corresponding scale gives the estimate of $\eta$.

To assess recovery at nominal and displaced values of both parameters, the evaluation spans nine truth points,
\begin{equation}
 \Theta_{\rm eval}=\{-0.6,0,0.6\}\times\{-0.025,0,0.025\}.
\end{equation}
A fitted coupling can differ from the generating value because of bias and statistical fluctuations across datasets. At each parameter setting, the root-mean-square error (RMSE) measures their combined effect in coupling units. We give the nine settings equal weight by averaging their RMSEs. For network $m$ this gives
\begin{equation}
 R_m=\frac{1}{J}\sum_{j=1}^{J}
 \left[\frac{1}{E}\sum_{b=1}^{E}
    \left(\widehat c_{mjb}-c_j\right)^2\right]^{1/2},
 \qquad J=9 .
 \label{eq:rmse}
\end{equation}
Here $\widehat c_{mjb}$ is the estimate from mock dataset $b$ at generating setting $j$, with $E$ datasets per setting. The mean across $M$ networks is $\overline R=M^{-1}\sum_m R_m$, and $g_m=R_m^{\rm orig}-R_m^{\rm repair}$ gives the gain from the edit. We also compute the bias in the coupling and jet energy scale at each setting. We average the absolute biases so that shifts of opposite sign cannot cancel across settings.

\subsection{How a systematic tilt in the likelihood shifts the fit}

Suppose we repeat the measurement with the same trained estimator but collect more collision events in each experiment. The fitted coupling will generally fluctuate less from one experiment to the next. A systematic error in the learned likelihood behaves differently. If, near the true coupling, the network gives slightly too much weight to larger values, further events reinforce that preference on average. The fitted values can then cluster more tightly around a displaced maximum. How far the maximum moves depends on how sharply the data constrain the parameters in the direction of the tilt.

We can quantify the tilt by how the event log likelihood changes as we vary the physics parameters $\theta=(c,\eta)$ with the network weights $\phi$ held fixed. This slope is the score $s_\phi(x;\theta)=\nabla_\theta\log r_\phi(x;\theta)$. When proposing an edit we differentiate a different quantity, the training loss, with respect to the weights $\phi$. At the generating point $\theta_0$, the score of an exact regular likelihood has zero mean. The learned likelihood can have a residual mean score $\bar s_\phi=\E_{\theta_0}[s_\phi(X;\theta_0)]$.

Let $\theta^\star_{\phi,n}$ denote the parameter values that maximise the expected log likelihood per event, $\E_{\theta_0}[\ell_\phi(\theta)/n]$, averaging over collision data and the auxiliary measurement at fixed weights and reference sample. At a nearby interior maximum the expected slope vanishes. Expanding this condition to first order about $\theta_0$ relates the displacement $\delta=\theta^\star_{\phi,n}-\theta_0$ to the local curvature,
\begin{equation}
 \left[J_\phi+
 \begin{pmatrix}0&0\\0&(n\sigma_a^2)^{-1}\end{pmatrix}\right]\delta
 \simeq \bar s_\phi,\qquad
 J_\phi=-\E_{\theta_0}[\nabla_\theta s_\phi(X;\theta_0)] .
 \label{eq:pseudotrue}
\end{equation}
For a single parameter, equation~\eqref{eq:pseudotrue} amounts to dividing the residual slope by the curvature. A shallow peak moves further under the same tilt. In the full log likelihood both the mean collision score and the collision curvature grow in proportion to $n$, so the resulting displacement can persist as more events are collected. The auxiliary measurement contributes once per dataset and adds curvature only along the jet-energy-scale direction. The off-diagonal entries of $J_\phi$ couple that direction to the fitted coupling, so an error in the learned scale response can also move the coupling estimate. The expansion assumes a differentiable likelihood and nonsingular combined curvature near the maximum.

The curvature thus tells us which combinations of parameters the measurement constrains well, a connection also used in information-geometric studies of collider measurements~\citep{brehmer2017geometry}. An edit that reduces the tilt in a weakly constrained direction can matter to the fitted coupling even if it changes the average training loss very little. The paired fits test whether the edit delivers that improvement on datasets of the size used in the analysis.

\subsection{Assessing uncertainty separately from point reconstruction}

Suppose a weight edit moves the peak of the profile likelihood closer to the true coupling and makes the profile fall away more steeply. The point estimate improves for that dataset, and at a given cutoff the steeper profile can give a shorter confidence set. At the same coverage, a shorter set gives a more precise measurement. An apparent gain in precision is misleading if the sets miss the true coupling too often. We therefore assess coverage alongside the coupling RMSE. We use one set of mock datasets to calibrate each estimator's cutoff and a separate set to measure how often the resulting confidence sets contain the generating coupling, following the logic of frequentist likelihood-free inference~\citep{dalmasso2020confidence,dalmasso2024lf2i}.

For a mock or observed dataset $D$, the profiled likelihood-ratio statistic measures the loss in fit when $c$ is fixed and $\eta$ is allowed to adjust,
\begin{equation}
 q_\phi(c;D)=2\left[\max_{c',\eta}\ell_\phi(c',\eta)
                  -\max_{\eta}\ell_\phi(c,\eta)\right].
\end{equation}
At nominal coverage $\alpha$, the confidence set contains every coupling for which this loss is within the calibrated threshold $t_{\phi,\alpha}(c)$,
\[
 C_{\phi,\alpha}(D)=\bigl\{c\in[-1,1]\mid q_\phi(c;D)\leq t_{\phi,\alpha}(c)\bigr\}.
\]
The true jet energy scale is unknown when we analyse data, yet it affects the distribution of $q_\phi$ even at a fixed coupling. A cutoff appropriate to the nominal scale could reject the true coupling too often at a shifted scale. To choose a cutoff that allows for this uncertainty, we assemble mock datasets at each trial coupling $c$ for three scale shifts, $\eta=-0.025,0,0.025$. For each shift we take the higher empirical $\alpha$-quantile of $q_\phi(c;D)$, with $\alpha=0.68$ or $0.95$. The largest of the three quantiles defines $t_{\phi,\alpha}(c)$. We repeat the calibration for each estimator because an edit can change the distribution of the statistic.

This choice makes the confidence sets conservative. At a scale where a smaller cutoff would suffice, a mock dataset can give a statistic above its own scale-specific cutoff but below the common one. The set then retains the true coupling, so coverage at that scale can exceed the nominal level. This helps explain the overcoverage observed before and after repair in Figure~\ref{fig:coverage}. We test coverage at the three simulated scale settings, and the finite calibration samples introduce Monte Carlo uncertainty in the thresholds.

In the final comparisons, we use $K=128$ mock datasets at each coupling and nuisance setting to calibrate the cutoff and $E=256$ per generating setting to evaluate coverage and coupling recovery. These draws come from separate samples of simulated events. Appendix~\ref{app:fit-grid} gives the numerical procedure for the fit and the confidence-set endpoints. Increasing $K$ and $E$ reduces fluctuations in the repeated draws. Section~\ref{sec:source-sensitivity} examines how the results depend on the original simulated events themselves.

\section{Collider benchmark and neural estimator}
\label{sec:collider}
\subsection{Separating known amplitude structure from detector response}

In electroweak Higgs production with two jets, the amplitude fixes the quadratic dependence of the unnormalised rate on the CP-odd coupling. The interference term can have either sign, but its size is bounded because the rate must remain nonnegative. Both properties survive showering, detector response and selection. By building them into the estimator before training, we can test whether repair improves a coupling measurement that already makes use of the known production physics.

In simulated proton--proton collisions at $\sqrt{s}=13$ TeV, a single dimensionless coefficient $c$ controls the CP-odd $HZZ$ and $HWW$ production interactions,
\begin{equation}
 \mathcal L_{\rm odd}=
 -\frac{c}{4\Lambda}H Z_{\mu\nu}\widetilde Z^{\mu\nu}
 -\frac{c}{2\Lambda}H W^+_{\mu\nu}\widetilde W^{-\,\mu\nu},
 \qquad \Lambda=1~{\rm TeV}.
 \label{eq:interaction}
\end{equation}
Here $\widetilde V^{\mu\nu}=\tfrac12\epsilon^{\mu\nu\rho\sigma}V_{\rho\sigma}$. The generator retains at most one insertion of the effective Higgs interaction per diagram. With the Standard Model-like term held fixed, this gives the production amplitude $\mathcal M=\mathcal M_0+c\mathcal M_1$. We keep its full square, which provides both the interference term linear in $c$ and the quadratic contribution from the CP-odd amplitude. The signal-only sample uses the same $H\to\gamma\gamma$ decay for every $c$, with Higgs mass 125 GeV and width 0.00407 GeV. Fixing the decay focuses the test on how this single phenomenological coupling changes the production kinematics.

The fit sees the reconstructed observables $x$, while the amplitude depends on the underlying partonic state $z$. For a given $z$, the detector response and selection at jet energy scale $\eta$ are described by the same nonnegative kernel $K_\eta(x,z)$ at every coupling. Integrating over the states that can give rise to $x$ therefore preserves the quadratic dependence of the unnormalised selected differential density,
\begin{equation}
 f(x;c,\eta)=\int \dd z\,K_\eta(x,z)
       |\mathcal M_0(z)+c\mathcal M_1(z)|^2
       = f_0(x;\eta)+c f_1(x;\eta)+c^2 f_2(x;\eta).
\end{equation}
The integration measure includes the positive leading-order sampling and parton-density factors. Where $f_0(x;\eta)>0$, define $u_\eta=f_1/f_0$ and $v_\eta=f_2/f_0$. Cauchy--Schwarz applied to this positive measure bounds the interference by the two squared-amplitude terms and gives
\begin{equation}
 v_\eta\geq0,\qquad u_\eta^2\leq4v_\eta .
 \label{eq:positivity}
\end{equation}
Dividing by the unnormalised density at zero coupling and the same jet energy scale gives $f(x;c,\eta)/f_0(x;\eta)=1+cu_\eta(x)+c^2v_\eta(x)$. For a measured $x$, the network thus needs two coefficients to describe the coupling dependence. It also needs the change in density at zero coupling as the jet energy scale moves away from its nominal value, since the reference in the fit is fixed at $c=\eta=0$. This reference must have support wherever the fit assigns a positive event density.

\subsection{Observables, selection and nuisance response}

We select a diphoton pair in the Higgs mass window and two jets with large dijet invariant mass and rapidity separation. Appendix~\ref{app:simulation-settings} gives the generator settings and selection cuts. Since the fit conditions on the event count, its sensitivity comes from the kinematic distributions of these selected events. Tagging-jet angles probe the tensor and CP structure of the interaction \citep{plehn2002structure,hankele2006anomalous}, and correlations with the surrounding phase space help constrain CP-violating Higgs--gauge interactions \citep{bernlochner2019angles}. Related studies exploit hadronic event shapes \citep{englert2012eventshapes} and the kinematics of gluon-fusion $Hjj$ with $H\to\tau\tau$ \citep{dolan2014cpgluon}.

The signed azimuthal separation of the jets retains information about interference between CP-even and CP-odd amplitudes \citep{hankele2006anomalous}. To preserve the sign, we subtract the azimuth of the backward jet from that of the forward jet, ordering the pair by rapidity for this purpose. Its sine and cosine preserve the orientation and are continuous across the angular wrap. The momenta and rapidity separation give the estimator information about the rest of the event, allowing it to learn how the angular distribution varies with the jet and diphoton kinematics. The eight inputs before standardisation are
\begin{align}
 x=\big(&\log[p_{\mathrm{T},j_1}/(30~{\rm GeV})],
 \log[p_{\mathrm{T},j_2}/(30\,{\rm GeV})],
 \log[m_{jj}/(400\,{\rm GeV})],|\Delta y_{jj}|,\nonumber\\
 &\sin\Delta\phi_{jj}^{\rm signed},
 \cos\Delta\phi_{jj}^{\rm signed},\nonumber\\
 &
 \log[(p_{\mathrm{T},\gamma\gamma}+1\,{\rm GeV})/(100\,{\rm GeV})],
 (m_{\gamma\gamma}-125\,{\rm GeV})/(10~{\rm GeV})\big).
\end{align}

To model a common jet-energy-scale shift, we multiply both reconstructed jet four-vectors by $1+\eta$. Raising the scale can move a jet above its $p_T$ threshold or a jet pair above the $m_{jj}$ cut. We therefore recalculate the observables and reapply the selection at each nuisance value. The fit uses the resulting distribution of accepted events, with the auxiliary measurement in equation~\eqref{eq:likelihood} constraining $\eta$.

\subsection{Learning the coefficient functions}

Near $c=0$, the sign of the interference coefficient $u_\eta(x)$ tells us whether a small positive coupling raises or lowers the differential rate at the measured kinematics $x$. The network learns this coefficient, the quadratic coefficient $v_\eta(x)$ and the change in the zero-coupling density with the jet energy scale. We use two tanh hidden layers of width $w$, with architecture $9\to w\to w\to3$. The nine inputs are the eight standardised observables and $\eta/0.05$. The three outputs $(a_\phi,b_\phi,h_\phi)$ determine the numerator of the fixed-reference ratio through
\begin{align}
 u_\phi &=a_\phi,\qquad
 v_\phi=\frac{a_\phi^2}{4}+\softplus(b_\phi),\\
 \widetilde r_\phi(x;c,\eta)
 &=\exp[(\eta/0.05)h_\phi(x,\eta)]
       [1+c\,u_\phi(x,\eta)+c^2v_\phi(x,\eta)] .
 \label{eq:structured}
\end{align}
With $\softplus(b)=\log(1+e^b)$, the polynomial is $(1+ca_\phi/2)^2+c^2\softplus(b_\phi)$ and is strictly positive for finite network outputs. The exponential estimates $p(x\mid 0,\eta)/p(x\mid 0,0)$, the change in the selected-event density due to the jet energy scale at zero coupling. Multiplying by the polynomial accounts for the coupling dependence of the unnormalised density. Equation~\eqref{eq:normalizer} normalises the full product to give the ratio used in the fixed-count fit. At $\eta=0$ the exponential equals one. The complete ratio is identically one at $c=\eta=0$.

Changing the trainable network weights $\phi$ adjusts the response to the reconstructed event while preserving both the quadratic numerator and the positive likelihood ratio. The more accurate coupling estimates reported in Section~\ref{sec:results} are obtained within these physical constraints.

The generator event weights supply the training labels for the coupling coefficients. We generate collisions at $c=0$ and use the matrix element to determine how much each collision contributes at other couplings, reusing its shower and detector simulation. For each original event, the weights at $c=-1,0,1$ determine the polynomial $A_0+cA_1+c^2A_2$. Additional weights at $c=\pm0.5$ and $\pm1.5$ check this dependence. The resulting labels $u_J=A_1/A_0$ and $v_J=A_2/A_0$ describe the interference and squared CP-odd contributions for that generator state.

Different partonic configurations can produce the same measured jet and photon kinematics. The coefficients appropriate to a measured $x$ are therefore averages over the generator states compatible with $x$. At fixed $\eta$, weighting selected events in proportion to $A_0/w_{\rm sampling}$, where $w_{\rm sampling}$ is the original sampling weight, gives the reference distribution at zero coupling. Under this distribution the conditional means of $u_J$ and $v_J$ given $x$ are $u_\eta(x)$ and $v_\eta(x)$. We estimate these means by weighted squared-error regression on the event labels~\citep{brehmer2020mining,brehmer2020madminer}.

For the nuisance factor, a binary classifier learns to distinguish shifted from nominal events at $c=0$. At each training value of $\eta$, the two classes receive equal total weight and the same conditioning value $\eta$. Bayes' rule then gives class odds at a measured $x$ equal to $p(x\mid0,\eta)/p(x\mid0,0)$. The classifier logit, $(\eta/0.05)h_\phi(x,\eta)$, estimates the logarithm of these odds. Training spans five scale shifts from $-0.05$ to $0.05$ in steps of $0.025$. Appendix~\ref{app:training} specifies the training loss and the calculation of its weights, feature standardisation and target scales.

Widths $w=32,64,128$ give 1475, 4995 and 18179 trainable parameters. The width-128 networks let us test the repair on a larger model than those used during development while retaining the same architecture family. We train with Adam \citep{kingma2015adam}, using a learning rate of $10^{-3}$, batches of 1024 augmented rows, weight decay $10^{-5}$ and a limit of 160 epochs. To select the starting network by its fit to held-out simulation data, we check the unregularised validation loss every four epochs and replace the retained checkpoint when that loss falls by more than $10^{-7}$. Training stops after 12 checks without such an improvement or when it reaches the epoch limit.

\subsection{From simulated events to repeated fits}
\label{sec:event-samples}

We count each original simulated collision once. Its reconstructions at different jet energy scales share an event identifier and stay together when we divide the simulation into samples. A mock dataset represents one repetition of the measurement. For a full fit at a given $(c,\eta)$, we draw 1000 accepted events with replacement at that scale according to their coupling weights and add one auxiliary scale measurement. The same original collision can thus contribute to many mock datasets.

All 18 networks in the final test share the same original-event samples and start from initialisations reserved for this test, six at each width. The training sample contains 20000 events passing nominal selection, and another 2500 nominally selected events form the validation sample used to choose the checkpoint. The adaptation sample supplies a further 2500 events passing nominal selection. We split the original collisions in that sample before constructing the jet-energy-scale variations, so an event that crosses a cut stays with its other reconstructions. One part supplies the gradient for proposing weight changes and the other supplies the mock datasets for screening them.

The profiled fits use three further samples. The normalisation sample sets the scale of each candidate's predicted likelihood ratio. From the calibration sample we draw mock datasets to set its confidence-set thresholds. From the evaluation sample we draw mock datasets to measure coupling error and coverage. The three samples contain 2216, 2264 and 2242 events passing nominal selection, respectively. After selection we fix the edited weights and draw new calibration and evaluation datasets from these same samples in each of two runs. Table~\ref{tab:pools} gives the generated and selected event counts for all six roles.

\section{Applying VEFR to the Higgs benchmark}
\label{sec:method}
A useful repair must improve the estimate of the Higgs coupling when the jet energy scale is allowed to adjust in the fit. During development, our first search built an archive from prescribed changes to different parts of the network, including its shared hidden layers. In an early trial it improved 12 of the 18 networks tested. Varying the weights feeding the two coupling outputs together revealed useful corrections that the archive's proposal rules had missed. We therefore kept the search across layers and added a second search confined to the coupling-output weights and biases. This gives us a way to test whether the hidden layers already encode the information needed for a better coupling estimate.

Both searches screen their proposals with a fit at the nominal jet energy scale. The ranking can change when the jet energy scale is allowed to vary. We therefore reserve a place for the best qualifying output-only candidate alongside the archive candidates in both rounds of the more expensive profiled fit. We fixed this design before testing it on the separate network initialisations reported in Section~\ref{sec:results}. Appendix~\ref{app:training} gives the development sequence.

\subsection{Screening edits at the nominal jet energy scale}
Both searches use the gradient of the supervised loss in equation~\eqref{eq:loss}, evaluated at the original weights $\phi_0$ on the labelled part of the adaptation sample (Section~\ref{sec:event-samples}). We include the regression and classification terms and omit weight decay. The archive uses this gradient to construct edits, while the coupling-output search uses it to identify sensitive coordinates.

At $\eta=0$ the screening fit only scans the coupling. We use 41 grid points from $-1$ to $1$ and three generating couplings, $-0.6,0,0.6$. For each generating value we reweight nominal events in the screening sample using the simulator-labelled polynomial and draw 64 mock datasets of 200 events with replacement. Each candidate fits the same draws, with its polynomial normalised on the screening sample. For the screening error $R_{\rm proxy}$ we pool the squared errors over all draws and parameter points before taking the square root. The full fit profiles $\eta$ and averages the RMSEs across the tested settings as in equation~\eqref{eq:rmse}. Both choices can affect the ranking of candidates.

An edit with a small screening error can still distort the predicted dependence on the coupling. We monitor this dependence with $S_{\rm proxy}$, which compares the logarithms of the normalised coupling polynomials across the coupling grid. One polynomial comes from the edited estimator at the measured $x$, and the other comes from the matrix-element weights for the underlying generator state $z$. The detector-level ratio averages the joint ratio over generator states at fixed $x$. Taking the logarithm before this average changes the target, since in general
\begin{equation}
 \E[\log r_{\rm joint}(x,z;\theta)\mid x]
 \ne \log\E[r_{\rm joint}(x,z;\theta)\mid x]
\end{equation}
under the joint reference measure. We use $S_{\rm proxy}$ to limit how far agreement with these generator labels can deteriorate during screening. Paired fits subsequently determine whether the edit improves the measurement. Appendix~\ref{app:proxy-surface} gives the definition of $S_{\rm proxy}$.

An archive edit to a shared layer can also change the predicted response to a shift in jet energy scale. At $\eta=0$ the nuisance factor in equation~\eqref{eq:structured} equals one for every $h_\phi$, so candidates with different slopes can give the same nominal fit. The slope of the log nuisance factor before normalisation is $h_\phi(x,0)/0.05$ at that point. We therefore limit the unweighted root-mean-square change in $h_\phi(x,0)$ to 0.01 on the nominal screening sample. Archive candidates must also have finite screening scores and satisfy $S_{\rm proxy}(\phi)\leq 1.015\,S_{\rm proxy}(\phi_0)$.

\subsection{Building an archive of edits across layers}
Each archive proposal applies one to four elementary weight changes, called primitives, starting at $\phi_0$. A sparse primitive selects up to 16 weights with large gradient components in one matrix. A hidden-unit primitive changes a second-layer unit's incoming weights and outgoing coupling weights together. A rank-one or rank-two primitive uses leading singular vectors of the layer gradient. We also try shifts of a coupling-output bias and rescalings of a second-layer unit's incoming weights and bias with reciprocal changes to its outgoing coupling weights. These rescalings can change the function since $\tanh(fz)/f\ne\tanh(z)$ in general. For sparse and low-rank primitives in the output layer, we mask the nuisance column of the gradient.

We sample 2688 compositions per network using the schedule in Appendix~\ref{app:repair-settings}. To keep several kinds of edits available for the profiled fit, we group similar proposals in an archive, following MAP-Elites~\citep{mouret2015illuminating,cully2018quality,fontaine2020covariance}. The grouping uses the type and layer of the final primitive, the number of primitives, and bins of edit size, displacement norm, discrepancy and nuisance-output change. Within each group we retain the candidate minimising
\begin{equation}
 Q(\phi)=R_{\rm proxy}(\phi)+0.15\,S_{\rm proxy}(\phi)
               +0.01\,\|\phi-\phi_0\|_2.
\end{equation}
This score favours low screening error, agreement with the simulator labels and small displacements. From the retained candidates, we shortlist one for each combination of final change type and layer, with at most ten combinations. We choose each representative by screening RMSE, breaking ties by the discrepancy and then by displacement norm.

\subsection{Joint changes to the coupling outputs}
With the hidden layers and the nuisance output held fixed, the coupling-output search changes how the learned features contribute to the two coupling coefficients in equation~\eqref{eq:structured}. Among the weights and biases feeding these outputs, we select the 12 coordinates with the largest absolute gradient. We draw joint Gaussian displacements on these coordinates so that several output parameters can change together.

The search generates 512 proposals. We clip each displacement and score the resulting candidate $\phi$ by its coupling error with a penalty for changing the predicted coupling dependence,
\[
 R_{\rm proxy}(\phi)+10\max\{0,S_{\rm proxy}(\phi)/S_{\rm proxy}(\phi_0)-1.01\}.
\]
Among tied scores we favour smaller displacements. The scores of the clipped proposals determine which draws advance the search. We average the eight best draws before clipping to update the Gaussian mean and shrink its scalar standard deviation on a fixed schedule. This uses a simpler update than covariance adaptation or natural-gradient evolution strategies~\citep{hansen2016cma,wierstra2014natural}. Appendix~\ref{app:repair-settings} gives the settings.

To choose the coupling-output representative, we require finite scores, lower screening RMSE than the original network and $S_{\rm proxy}(\phi)\leq 1.01\,S_{\rm proxy}(\phi_0)$. We select the qualifying proposal with the lowest unpenalised screening RMSE, breaking ties by the discrepancy and then by displacement norm. If no proposal qualifies, this branch returns the original weights.

\subsection{Selecting a repair with the profiled fit}
The first round runs the profiled fit for the original network, up to ten archive candidates and the coupling-output representative. We use $K=32$ calibration and $E=32$ evaluation datasets per setting. We rank candidates by $L_q$ from equation~\eqref{eq:lcb}, using the squared error in the coupling for $L_{\rm task}$. Ties are broken by the gain in mean RMSE and then by candidate identifier. Up to three archive candidates advance alongside the coupling-output representative to the final comparison, with $K=128$ and $E=256$. These two selection rounds share a random seed, so their draws partially overlap.

In the final comparison we require $L_q>0$ and a reduction in the mean coupling RMSE in equation~\eqref{eq:rmse}. A gain in the average squared error can occur with an increase in the average of the individual RMSEs, so both conditions matter. We select the highest-ranked candidate that passes both checks and use the original weights if there is no qualifying edit. The decision uses the point estimates. The calibration draws let us assess the confidence sets. We fix the chosen weights before the two evaluations in Section~\ref{sec:evaluation-protocol}.

\subsection{Comparison with a search along one gradient direction}
A single further step along the training gradient might itself improve the coupling fit. To measure how much such a step can achieve, we evaluate the gradient of the supervised loss for the coupling-output weights and biases at $\phi_0$. We try seven step sizes along the resulting descent direction, starting each trial from the original network. Candidates need finite screening scores and $S_{\rm proxy}(\phi)\leq 1.015\,S_{\rm proxy}(\phi_0)$. Up to two candidates enter the first profiled comparison in order of screening RMSE, with ties broken by the discrepancy and then by step size. The best of these enters the final comparison and must pass the same acceptance rule as a \vefr{} edit.

The two procedures use the same simulated events (Section~\ref{sec:event-samples}) and the same number of mock datasets per candidate at each round of paired fits. Their computational budgets differ. \vefr{} screens 2688 archive proposals and 512 coupling-output proposals per network and can take more candidates into the profiled fits. Any difference in coupling accuracy must therefore be interpreted in light of both the directions explored and the additional computation available to \vefr{}. Within each procedure the proposal counts are the same at all network widths, although the cost of a forward pass, a gradient or a matrix decomposition depends on the width.

\section{Results}
\label{sec:results}
\subsection{Coupling recovery after fixing the weights}
\label{sec:evaluation-protocol}
\begin{figure}[!htb]
\centering
\includegraphics[width=\textwidth]{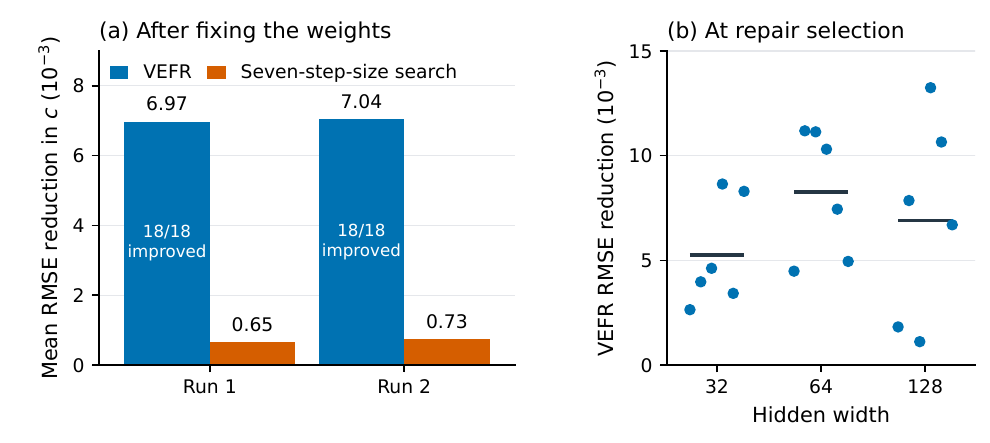}
\caption{Coupling RMSE reduction relative to the original networks. Panel (a) compares \vefr{} with one additional gradient step in two evaluations after fixing the weights. The bars give the mean gain across 18 networks on new mock datasets drawn from the same simulated events. Panel (b) shows \vefr{} gains on the datasets used to select the edits. Each dot is one network, and each horizontal line is the mean for that width.}
\label{fig:networks}
\end{figure}

After the repairs were fixed, all 18 networks recovered the Higgs coupling more accurately in each of two evaluations. The mean reduction in coupling RMSE relative to the original networks was $0.00697$ in the first run and $0.00704$ in the second (Figure~\ref{fig:networks}a). The 18 networks were trained on the same events, with six initialisations reserved for this test at each hidden width of 32, 64 and 128. The gain thus extends to a wider network than those used to develop the search.

An edit chosen for its performance on a set of mock experiments may owe some of its apparent gain to fluctuations in those particular datasets. For each evaluation we therefore kept the selected weights fixed and used a new random seed to draw mock datasets from the same calibration and evaluation event samples (Section~\ref{sec:event-samples}). Within each mock experiment all estimators fitted the same collision events and auxiliary measurement. The close agreement between runs shows that the gain survives a fresh draw of mock experiments from these events.

A single additional step along the training gradient also reduced the coupling RMSE, by about $0.0007$ on average in each run (Table~\ref{tab:results}). For this comparison we tried seven step sizes along one direction. The larger gain with \vefr{} was obtained with a wider choice of directions and a larger budget for testing candidates in the full fit.

\begin{table}[!htbp]
\caption{Reduction in coupling RMSE relative to the original networks. The reductions are averaged over 18 networks. Runs 1 and 2 test fixed weights on newly assembled datasets. The last row uses the datasets on which the edits were selected.}
\label{tab:results}
\centering
\begin{tabular}{lrrr}
\toprule
Stage & Networks improved by \vefr & \vefr{} gain & Seven-step-size gain\\
\midrule
Fixed weights, run 1 & 18/18 & 0.006970 & 0.000650\\
Fixed weights, run 2 & 18/18 & 0.007040 & 0.000732\\
\midrule
Repair selection & 18/18 & 0.006804 & 0.000608\\
\bottomrule
\end{tabular}
\end{table}

The full fit selected the coupling-output candidate for eight of the 18 networks. For these networks, changing the weights and biases of the two outputs was enough to give a more accurate coupling estimate from the information already present in the hidden layers. The other ten selected repairs came from the archive search. On the datasets used to select the edits, \vefr{} reduced the mean RMSE from 0.179663 to 0.172859 (3.79\%). The mean gains at widths 32, 64 and 128 were 0.005266, 0.008250 and 0.006897, respectively (Figure~\ref{fig:networks}b). The seven-step-size search reached 0.179055 at selection, with qualifying edits for five networks and the original weights retained for the other 13.

\subsection{Sensitivity to the original simulated collisions}
\label{sec:source-sensitivity}
Both fixed-weight evaluations draw on the same original collisions. If only a few simulated events populate a region of phase space, those events can recur in many mock experiments. To see how much the gain depends on the mix of collisions available, we resampled the original collisions separately within the normalisation, calibration and evaluation samples and repeated the fits with the weights fixed. Each collision and all its jet-energy-scale variations received the same multiplicity.

The mean gain across the 18 networks was positive in 31 of the 32 event resamples, and the average gain over all resamples was $0.007121$. In the remaining resample only six networks improved and the mean gain reversed sign, to $-0.002905$. Thus, even when the repair helps in repeated mock measurements, a change in the mix of simulated collisions can make the repaired networks perform worse on average. With 32 resamples, the frequency and size of such losses in other event samples remain poorly determined (Figure~\ref{fig:source} in Appendix~\ref{app:resampling-tail}). In the negative resample the mean coverage of the nominal 68\% sets fell by 1.74 percentage points after repair. Across all resamples, the largest increase in mean absolute jet-energy-scale bias was $4.36\times10^{-5}$.

On six prespecified networks we also checked how accurately the numerical fits resolve the gain. Doubling the number of mock datasets or refining the fit grid changed the mean gain by about $0.0002$, compared with a gain of about $0.0046$ on the finer grid at the larger draw count. All six networks improved in that comparison (Table~\ref{tab:numerical-resolution}).

\subsection{Coupling bias and conservative coverage}
\begin{figure}[!htb]
\centering
\includegraphics[width=\textwidth]{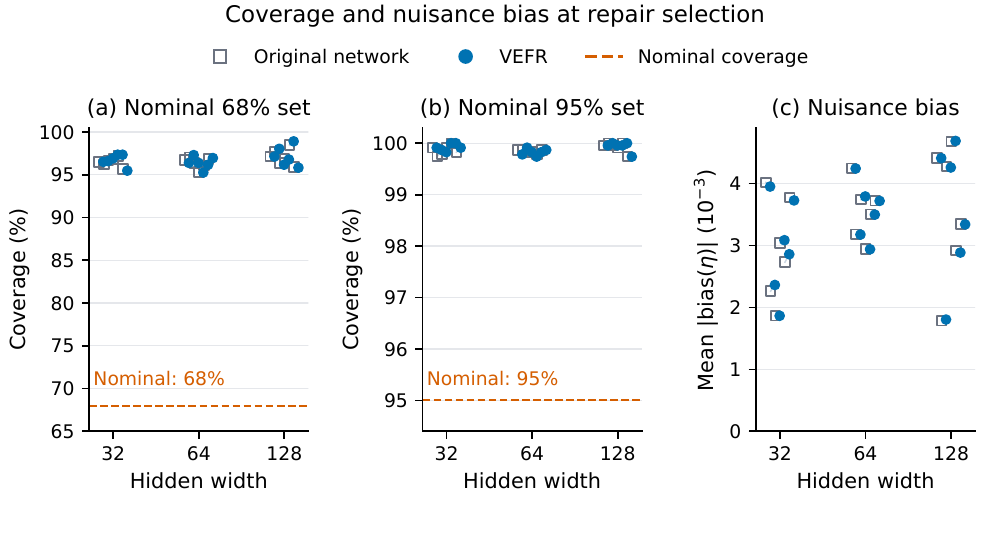}
\caption{Empirical coverage and mean absolute jet-energy-scale bias at repair selection, grouped by network width. Open squares denote the original network and filled circles the \vefr{} repair. Dashed lines show nominal coverage. We take the absolute bias at each generating setting before averaging.}
\label{fig:coverage}
\end{figure}

At repair selection, the nominal 68\% confidence sets contain the true coupling in about 97\% of the mock datasets on average, before and after repair. The nominal 95\% sets cover about 99.9\% (Figure~\ref{fig:coverage}). Thus the confidence sets include the true coupling more often than their nominal levels imply. The calibration uses the largest cutoff over the tested jet energy scales, which gives more inclusive sets wherever a smaller cutoff would have sufficed. The demonstrated gain concerns the accuracy of the fitted coupling. Comparing the precision of the confidence sets would require their lengths at matched coverage.

On the selection datasets, the mean absolute coupling bias falls from 0.086903 to 0.080457. The mean absolute bias in the jet energy scale moves from 0.003358 to 0.003367 at selection and increases by about $10^{-6}$ in each evaluation of the fixed weights. These are averages of the absolute bias across generating settings and networks.

\FloatBarrier
\section{Summary and Conclusions}
\label{sec:discussion}
Generating enough simulated events to learn an accurate likelihood can account for a large part of the cost of an SBI analysis. Once the estimator is trained, the available simulations still contain information about the measurement itself. At known generating parameters we can ask whether a change to the network improves the fit to a dataset of the size used in the analysis. In \vefr{}, gradients guide the search over changes to the trained weights, and fits with the original and edited estimators on the same mock datasets determine which change to keep. For fixed inputs and random seeds, the numerical procedure is deterministic and runs without language models or agents.

In the Higgs application, all 18 networks gave more accurate estimates of the coupling after repair. The mean reduction in coupling RMSE was about $0.007$ in each evaluation of the fixed weights (Table~\ref{tab:results}). Both runs used new mock datasets drawn from the original simulation samples. Eight selected edits changed only the coupling outputs, so for these networks the learned representation already supported a more accurate fit. We selected the edits by coupling error. The confidence sets covered the true coupling more often than their nominal levels both before and after repair.

To apply \vefr{} in another analysis, we need access to the trained estimator's weights and a training loss whose gradient we can evaluate on simulated data. Those simulations must cover the parameter values and nuisance settings relevant to the measurement, with enough events to assemble representative mock datasets. The analysis also needs a repeatable fitting procedure and a measure of what counts as an improvement. Running the original and edited estimators on the same datasets lets us compare their errors against the same generating values. Once an edit is chosen, a separate set of mock datasets provides an evaluation of the fixed weights.

Other SBI analyses can use weight repair after training because the fit itself determines whether an edit helps. This decision requires a way to measure the error on simulations and can be made independently of any diagnosis of its cause. The improvement we obtained at a fixed simulation budget makes it plausible that repair can bring an estimator trained on fewer simulated events to the accuracy required by the measurement.

\bibliography{references}
\bibliographystyle{tmlr}
\appendix
\section{Simulation, training and weight repair in the Higgs benchmark}
\label{app:training}
Each simulated collision carries generator weights for several values of the Higgs coupling. We keep these weights with its reconstructed jets and photons as we shift the jet energy scale and reapply the selection. In the training sample the weights supply labels for learning the coupling dependence from the reconstructed kinematics. To propose a repair, we compute the training gradient on a separate sample of labelled collisions. Other collisions supply the mock datasets used to screen the edits and to compare the original and repaired coupling fits. Table~\ref{tab:pools} gives the number of original collisions used for each purpose.

The supplied script \path{analysis/make_figures.py} reads the numerical summaries in \path{data/source/} to reproduce the reported gains and the results figures. The manifest \path{provenance/source_manifest.json} identifies these inputs and the revision of the original analysis (\texttt{f8a59a63}). Repeating the full search requires the simulated events, training checkpoints, search code and configuration held in the authors' private research archive. The configuration specifies the proposal sampling rules, archive bin boundaries and search seeds.

\subsection{Event generation and selection}
\label{app:simulation-settings}
We generate proton--proton collisions at $\sqrt{s}=13$~TeV with MadGraph5\_aMC@NLO~3.5.7~\citep{alwall2014madgraph} and the Higgs Characterisation model~\citep{artoisenet2013higgs,maltoni2013higgs}. The complete tree-level electroweak amplitude in the four-massless-quark-flavour scheme includes Higgs-strahlung and interference among the electroweak production contributions. Pythia~8 simulates the shower and Higgs decay~\citep{sjostrand2015pythia}, and Delphes models the detector response~\citep{deFavereau2014delphes}.

At generation we require partonic jets with $p_{\mathrm{T},j}>20$~GeV and pseudorapidity magnitude below 5, $m_{jj}>200$~GeV and a minimum jet pseudorapidity separation of 2. At detector level we use anti-$k_T$ jets of radius $R=0.6$. An event must contain at least two jets within pseudorapidity magnitude 4.5 and two photons within 2.5. Of the objects within these acceptances, we take the two highest-$p_{\mathrm T}$ jets and photons. Both selected jets must have $p_{\mathrm T}>30$~GeV and both photons $p_{\mathrm T}>25$~GeV. We also require $m_{jj}>400$~GeV, $|\Delta y_{jj}|>2.8$ and $105<m_{\gamma\gamma}<145$~GeV, and impose $\sqrt{\hat s}<1$~TeV using the incoming parton momenta.

\subsection{Supervision and training}
A training row contains the reconstructed observables of one selected event, a conditioning value of $\eta$, and the targets for one of the two training tasks. For example, an event that passes selection both at the nominal jet energy scale and at a shifted scale supplies two reconstructions of the same underlying collision. A regression row pairs a reconstruction with the generator coefficients $(u_J,v_J)$. For classification at a given $\eta$, we give reconstructions at that scale the label $y=1$ and nominal reconstructions the label $y=0$, with the same conditioning value for both classes.

\begin{samepage}
The row weights $\omega_{\rm coef}$ and $\omega_{\rm cls}$ select the relevant term in the loss. The classification weight is zero on a regression row, and the coefficient weight is zero on a classification row. With $N_{\rm rows}$ rows in the sum, the loss is
\begin{align}
 \mathcal L(\phi)=\frac{1}{N_{\rm rows}}\sum_i\bigg\{
 &\frac{\omega_{{\rm coef},i}}2
 \left[\frac{(u_\phi-u_{J,i})^2}{s_u^2}
       +\frac{(v_\phi-v_{J,i})^2}{s_v^2}\right]\nonumber\\
 &+\omega_{{\rm cls},i}
     [\log(1+e^{t_i})-y_i t_i]\bigg\}
 +\frac{\lambda}{2}\sum_{\ell}\|W_\ell\|_F^2 ,
 \qquad t_i=(\eta_i/0.05)h_\phi .
 \label{eq:loss}
\end{align}
\end{samepage}
At each of the five training nuisance values, we weight selected events in proportion to $A_0/w_{\rm sampling}$, where $w_{\rm sampling}$ is the original event's sampling weight. We normalise these weights within each nuisance value, give the five values equal weight, and give the two classes equal total weight in the classification term. We use the training events accepted at the nominal scale to standardise the observables. Reconstructions of the training sample at all five scales supply the target scales $s_u=0.233386$ and $s_v=0.033310$, which give the two regression errors comparable weight.

We exclude biases from the weight penalty and use Adam with $(\beta_1,\beta_2)=(0.9,0.999)$ and numerical stabiliser $10^{-8}$. The validation loss determines the checkpoint at which repair begins (Section~\ref{sec:collider}). Reusing events at the five scales and for the two training tasks gives 299482 training rows and 37426 validation rows. Table~\ref{tab:pools} counts the original collisions used to construct the samples. For training, validation and adaptation we take the initial portion of each generated sample that supplies the specified number of events passing the nominal selection. All scale variations for that role are built from this same portion, which includes events that can enter the selection after rescaling. For normalisation, calibration and evaluation we use the full generated samples, recorded as \texttt{wgr\_direct\_z\_20261107\_001}, \texttt{wgr\_direct\_k\_20261107\_001} and \texttt{wgr\_direct\_e\_20261107\_001}, respectively.

\begin{table}[!hb]
\caption{Simulation budget by role. Each generated collision counts once. The final column counts events from the portion used that pass the reconstruction cuts at $\eta=0$.}
\label{tab:pools}
\centering
\begin{tabular}{lrrr}
\toprule
Role & Events generated & Events used before cuts & Selected at $\eta=0$\\
\midrule
Training & 180000 & 177468 & 20000\\
Validation & 25000 & 22150 & 2500\\
Adaptation & 25000 & 22381 & 2500\\
Normalisation & 20000 & 20000 & 2216\\
Calibration & 20000 & 20000 & 2264\\
Evaluation & 20000 & 20000 & 2242\\
\bottomrule
\end{tabular}
\end{table}

The initial development group comprised 30 networks, five for each combination of width (32 or 64) and number of selected training events (4000, 8000 or 20000). An earlier trial of the structured search improved 12 of the 18 networks tested with that version. We then tested the combined archive and coupling-output searches described in Section~\ref{sec:method} on 12 further development networks trained with 20000 selected events, six at each width with initialisation seeds 88701--88706. Seven selected edits came from the coupling-output search and five from the archive. With the procedure fixed, we applied it to the 18 networks reported in Section~\ref{sec:results}. These used fresh initialisation seeds 88901--88906 at widths 32 and 64 and 89001--89006 at width 128. All widths use the same Glorot initialisation rule.

\subsection{Edit directions and archive descriptors}
In the archive search, a primitive is one elementary change to the weights. We compute the gradient $g$ at the original weights and normalise the proposal directions so that a sampled step size controls the displacement. A sparse primitive selects the 1, 2, 4, 8 or 16 coordinates of a weight matrix with largest $|g_i|$. On those coordinates $S$, the step direction is $-g_S/(\operatorname{RMS}(g_S)+\epsilon)$, with a small positive stabiliser $\epsilon$. A hidden-unit primitive normalises the negative gradients of its incoming weights and outgoing coupling weights separately. For a rank-$r$ change in $W_1$ or $W_2$, the direction is $-U_rV_r^\top$ divided by its Frobenius norm, where $U_r$ and $V_r$ contain the leading $r$ singular vectors of the selected gradient matrix. Coupling-output biases receive signed scalar displacements. To rescale a second-layer unit, multiply its incoming weights and bias by one factor and divide its outgoing coupling weights by that factor.

The archive groups candidates so that different types of edit reach the profiled fit. Each group has a common final primitive type and layer, number of primitives, and bins for edit size, displacement norm, relative change in $S_{\rm proxy}$ and change in nuisance output. We assign the logarithmic edit-size bin using the sum of the counts for each primitive. A sparse primitive contributes its number of selected coordinates, a rank primitive contributes the number of matrix entries, and a hidden-unit change, rescaling or output-bias shift contributes one. An edit that revisits a coordinate counts it again. Within each group we retain the candidate with the smallest composite score $Q$. To shortlist candidates for the profiled fit, we order these representatives by screening RMSE, then by $S_{\rm proxy}$ and displacement norm, taking one candidate for each final type and layer pair.

\subsection{Search schedules and evaluation seeds}
\label{app:repair-settings}
For the structured archive, we generate 96 compositions in each of 28 rounds, giving 2688 candidates per network. We sample primitive scales log-uniformly between $10^{-4.5}$ and $10^{-0.7}$ and multiply the completed displacement by $\max(0.15,0.92^k)$ in round $k$. Every proposal starts from the original weights.

The coupling-output search uses a Gaussian displacement distribution with initial mean zero and standard deviation 0.025. In each of 16 generations we draw 32 displacement vectors and clip their coordinates to $[-0.12,0.12]$ when forming candidates. The scores of the clipped candidates identify the best eight draws. We set the next Gaussian mean to the average of the corresponding unclipped vectors and shrink the scalar standard deviation according to $\sigma\leftarrow\max(0.002,0.82\sigma)$.

For the gradient comparator, we RMS-normalise the negative gradients at the original weights of the coupling-output weights and biases separately. We apply the resulting direction with seven alternative single-step scales $0.0003,0.001,0.003,0.01,0.03,0.06,0.1$.

The first and final selection fits share a sampling seed and partially overlapping draws. The evaluations of the fixed weights use the seeds 2026111401 and 2026111402. The numerical check with twice as many draws uses the six prespecified network indices $0,5,6,11,12,17$.

\subsection{Discrepancy used to screen edits}
\label{app:proxy-surface}
For an event reconstructed as $x_i$ from the generator state $z_i$, we can compare the coupling dependence predicted by the network with that supplied by the simulator. At the nominal jet energy scale the network predicts $P_\phi(x_i;c)=1+c\,u_\phi(x_i,0)+c^2v_\phi(x_i,0)$. The generator labels give $P_J(z_i;c)=1+c\,u_{J,i}+c^2v_{J,i}$ for the underlying event. We normalise each polynomial at each coupling using the same screening sample, then average their squared log differences,
\begin{equation}
 S_{\rm proxy}(\phi)=\frac1{|\mathcal C|}
 \sum_{c\in\mathcal C}\sum_i\overline\omega_i
 \left[\log\frac{P_\phi(x_i;c)}
                  {\sum_k\overline\omega_kP_\phi(x_k;c)}
       -\log\frac{P_J(z_i;c)}
                  {\sum_k\overline\omega_kP_J(z_k;c)}
 \right]^2,
\end{equation}
where $\mathcal C$ is the 41-point coupling grid and $\overline\omega_i$ are the nominal coefficient weights normalised to sum to one. Bounding the increase in $S_{\rm proxy}$ during screening limits how far the edit can degrade agreement with the generator labels. Several generator states can produce the same measured $x_i$, so the gain in the detector-level measurement is assessed separately by the paired fits.

\section{Uncertainty calculations}
\label{app:uncertainty}
The estimated gain varies with the mock datasets used to measure it, the original simulated events available to construct those datasets and the trained networks included in the average. The paired standard error varies the mock datasets with the original simulations fixed. Resampling the original collisions tests the sensitivity of the fixed edits to the finite simulation samples used for inference. Resampling the network outcomes measures the variation in the mean over networks trained and evaluated on shared samples.

\subsection{Paired fits and resampling of the original events}
For a given edit and generating parameter point, the original and edited networks fit the same collision events and auxiliary measurement in each mock dataset. The sample variance of the paired squared-error differences gives the conditional standard error in equation~\eqref{eq:lcb}. The aggregation assumes independent draws of mock datasets across generating points conditional on the available simulations.

To vary the original events, we take 32 multinomial resamples of their identifiers. We resample separately within the normalisation, calibration and evaluation samples. If a collision occurs twice in one resample, it carries multiplicity two wherever it enters the comparison, for every jet energy scale, coupling and network. Events with zero multiplicity are omitted before normalisation. The network weights and chosen edits are held fixed throughout.

\subsection{Resolution of the resampling tail}
\label{app:resampling-tail}
For the 32 event resamples, the smallest mean gain is $g_{(1)}=-0.00290456$ and the second smallest is $g_{(2)}=0.00014747$. The estimated lower tail depends on how we locate the 2.5th percentile between these observations. The ordered gains span 31 intervals. At probability 0.025 the quantile position lies $31\times0.025=0.775$ of the way from the first gain to the second. The \texttt{higher} convention selects the upper neighbour $g_{(2)}$, while linear interpolation gives
\begin{equation}
 (1-0.775)g_{(1)}+0.775g_{(2)}=-0.00053924 .
\end{equation}

\begin{figure}[!htb]
\centering
\includegraphics[width=0.88\textwidth]{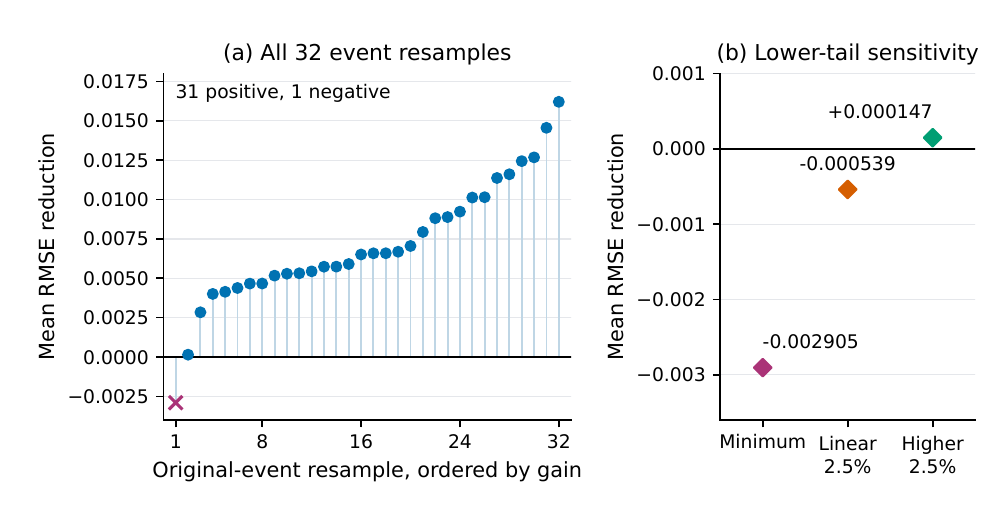}
\caption{Coupling RMSE reduction across 18 fixed repairs when the original events used for normalisation, calibration and evaluation are resampled. Panel (a) orders the 32 resamples by their mean gain. Panel (b) compares the minimum with the 2.5th percentiles obtained by linear interpolation and by taking the higher neighbour.}
\label{fig:source}
\end{figure}

The empirical 2.5th and 97.5th percentiles of the mean repair gain give ranges $[0.000147,0.016202]$ under the higher convention and $[-0.000539,0.014923]$ under linear interpolation. For the gain over the seven-step-size comparator, the linearly interpolated range is $[-0.001579,0.013644]$. The sign change at the lower endpoint limits what can be inferred about the extreme lower tail from these 32 resamples.

\subsection{Variability across trained networks}
For the network bootstrap we draw with replacement from the 18 network outcomes measured on the datasets used to choose the edits. Across 200000 resamples, the central 95\% bootstrap range for the mean selection-stage gain is $[0.005198,0.008425]$. The mean gain relative to the seven-step-size comparator is 0.006196. These quantities describe variability across the tested networks conditional on the shared simulation samples and the selection datasets.

\subsection{Numerical evaluation of the fit and confidence sets}
\label{app:fit-grid}
The final comparisons use a grid of $81\times41$ points in the coupling $c$ and jet energy scale $\eta$. To check the effect of the grid spacing, we repeat the fits on a nested $41\times21$ grid using the same mock datasets. At each coupling node we first find the maximum of the log likelihood over the nuisance grid. For an interior maximum we fit a local quadratic through the maximum and its two neighbours. We use its peak if the quadratic is concave and the peak lies between the neighbouring nodes. Otherwise we retain the grid maximum. The fitted coupling is the coupling node with the highest profiled log likelihood.

To find the confidence-set endpoints, we evaluate $q_\phi(c;D)-t_{\phi,\alpha}(c)$ on the coupling grid. Where the difference changes sign between adjacent nodes, we locate its zero by linear interpolation. The confidence set includes every connected component on which the interpolated difference is nonpositive, including components that extend to the boundary of the coupling range.

For the six prespecified networks, doubling the number of draws on the fine grid changes the mean gain by 0.000202 using the unrounded values (Table~\ref{tab:numerical-resolution}). At the doubled counts, refining the grid changes it by 0.000222. All six networks improve on the fine grid at the higher counts.

\begin{table}[!htb]
\caption{Mean reduction in coupling RMSE on six networks. For each grid we give the calibration count $K$ per coupling and nuisance setting and the evaluation count $E$ per generating setting.}
\label{tab:numerical-resolution}
\centering
\begin{tabular}{lrrr}
\toprule
Grid in $(c,\eta)$ & $K$ & $E$ & Mean gain\\
\midrule
$81\times41$ & 128 & 256 & 0.004814\\
$81\times41$ & 256 & 512 & 0.004613\\
$41\times21$ & 256 & 512 & 0.004835\\
\bottomrule
\end{tabular}
\end{table}

\end{document}